\documentclass[12pt]{iopart}

\usepackage{verbatim} 

\usepackage{graphicx}
\usepackage{float}
\restylefloat{figure}

\usepackage{amssymb,latexsym}
\usepackage{amsfonts}
\usepackage{amsthm}
\usepackage{stmaryrd}
\usepackage{color}

\newcommand{\id}{\textrm{d}}

 \let\be=\beta

\let\De=\Delta

\newcommand{\caI}{{\mathcal I}}

\newcommand{\caO}{{\mathcal O}}
\newcommand{\caP}{{\mathcal P}}

\newcommand{\caT}{{\mathcal T}}

\begin{document}
\newcommand{\pr}{\partial}
\newcommand{\p}{\prime}
\newcommand{\w}{\omega}
\newcommand{\lt}{\mathcal{L}}
\newcommand{\Fa}{F\left[\begin{array}{c|c} x & x^\p \\ t+\Delta t & t \end{array} \right]}
\newcommand{\Fb}{F\left[\begin{array}{c|c} x_t & x_0 \\ t & t_0 \end{array} \right]}
\newcommand{\ra}{\rangle}
\title{Dynamical fluctuations for periodically driven diffusions}
\author{Navinder Singh}
\address{Department of  Applied Mathematics, Faculty of Sciences, H.I.T.-- Holon Institute of
Technology, Holon, Israel}
\eads{\mailto{navinder.phy@gmail.com}}
\author{Bram Wynants}\address{Institute of Theoretical Physics, K.~U.~Leuven, Leuven, Belgium}
\eads{\mailto{Bram.Wynants@fys.kuleuven.be}} 

\begin{abstract}
We study dynamical fluctuations in overdamped diffusion processes
driven by time periodic forces. This is done by studying fluctuation functionals
(rate functions from large deviation theory), 
of fluctuations around the non-equilibrium steady regime. 
We identify a concept called traffic. 
This traffic, which was introduced in the context of 
non-equilibrium steady state statistics, is extended here for time-dependent 
but periodic forces.
 We discuss the fluctuation functionals of occupations and currents,
and work out some specific examples.
The connection between these and non-equilibrium thermodynamic potentials, their 
corresponding variational principles and their Legendre transforms, are also discussed.
\end{abstract}

\maketitle

\section{Introduction}

Fluctuation theory is a cornerstone of statistical mechanics. 
On the one hand
fluctuation theory naturally provides variational characterizations of the stationary regime.
On the other hand, the fluctuations themselves are very interesting,
especially because in recent years all kinds of mesoscopic systems have become
observable and manageable. Think of e.g. molecular motors and transport through nanotubes
in biophysics. Furthermore, and because of this, fluctuation theory has already helped in understanding
nonequilibrium statistical mechanics, see e.g. \cite{ons} - \cite{cmaes07a}.  \\

\noindent In fluctuation theory one one has to distinct between static  and dynamical fluctuations \cite{cmaes07a}.
In static fluctuation theory one conditions on having a stationary regime in the infinite
past, and measures the system at the present (time zero). This has been extensively
studied in \cite{jona1,jona2} for the hydrodynamic limit of several driven lattice gases.
There, a Hamilton-Jacobi equation is derived for the fluctuations in the spirit of \cite{kubo}.\\

\noindent We will concentrate on dynamical fluctuation theory, as started by Onsager and Machlup in 1953 \cite{ons}. 
In this theory, the system is continuously measured, and the fluctuation is a sequence of correlated 
non-equilibrium states (the time interval between two successive measurements is less 
than the correlation time). If the measurement time is infinite, the measured 
value of the observable will coincide with the `typical' value.  For finite times, averaged 
observables may or may not coincide with this 
typical value. If measurement times are large, then the fluctuations have
to persist for a long time to be significant: one enters in the regime of large deviations.\\

\noindent Extensions of the Onsager-Machlup theory have been made
in recent years to nonequilibrium dynamics. We will continue in the lines
of \cite{cmaes08a,cmaes08b,cmaes08c}.  
More precisely, this work is an extension of our previous work in \cite{cmaes08a}.
In \cite{cmaes08a} we have considered dynamical fluctuations of driven (by a time independent drive) overdamped diffusions.
The most important results there are:\\
1. The introduction of the concept of traffic, which
complements the concept of entropy.\\
2. The computation of the joint fluctuations of occupations and currents.\\
3. A formulation of dynamical fluctuations in terms of thermodynamical potentials.\\
4. An explanation of entropy production principles close to equilibrium.\\
In the present contribution we carry on the program of \cite{cmaes08a}
 to time-dependent (but periodic) dynamics like in \cite{nav}. We will
start here by introducing the model we work with.

\subsection{The model}
To introduce some notation, let us start with a simple model for an overdamped diffusion of a Brownian particle in $d$ dimensions.
Here and in the rest of this text we will always assume the process to be ergodic.
It is described by the Langevin equation (using the It\^o interpretation):
\begin{equation}\label{eqref}
dx_t = -\chi(x_t)\nabla U(x_t)dt + \nabla D(x_t)dt + \sqrt{2D(x_t)} dB_t
\end{equation}
Here, $x_t$ is a $d$-dimensional vector describing the position of the particle. The $d$-dimensional space that 
the particle can move in is denoted by $\Omega$. 
$B_t$ is a $d$-dimensional vector of standard Gaussian white noises, $\chi$ is the mobility matrix and $D$ is the diffusion matrix.
Using the Einstein relation we have $\chi = \beta D$, where $\beta$ is the inverse temperature of the surrounding fluid (bath). 
The potential $U$ can be seen as the energy of the system (particle). 
We will assume that $\Omega$
is either compact with periodic boundary conditions, or without boundaries. Think of e.g.
a $d$-dimensional torus or $\mathbb{R}^d$. In the latter case the potential $U$ should be sufficiently
confining, which makes sure that the particle does not `escape to infinity' (which happens e.g. for
pure diffusion with $U=0$ in $\mathbb{R}^d$). An example of such a confining potential
is a harmonic potential $U = \frac{k}{2}\sum_{i=1}^d x_i^2$.\\

\noindent The Langevin equation defines trajectories of the particle, a trajectory being the position
of the particle as a function of time.
Instead of considering trajectories, one can also take a probability density $\mu_0(x)$
as an initial condition, and see how this evolves in time. This evolution is given by
 the Fokker-Planck equation:
\begin{equation}\label{eqfokpla}
\frac{d\mu_t(x)}{dt}  = \nabla\cdot[D(x)\nabla\mu_t(x)+\chi(x)\nabla U(x)\mu_t(x)]
\end{equation}
The Boltzmann distribution
\[ \rho_0(x) = \frac{1}{Z}e^{-\beta U(x)} \]
solves (\ref{eqfokpla}) with left-hand side zero, and is thus a stationary distribution for this process.
We assume that $U$ and $\chi$ are such that this stationary distribution is unique and that 
for any $\mu_0$ we have that $\mu_t\to \rho_0$ for $t\to\infty$ (for more explanation and mathematical rigor, see \cite{kry}).
Indeed, we are dealing here with a (simple) equilibrium process. This process we will use as our basic reference process.\\

\noindent We add to the reference model (\ref{eqref}) a time-dependent force $f_t$. We take the
time-dependence to be periodic with period $\tau$: $f_t = f_{t+\tau}$. The Langevin equation of the new stochastic process
is now given by:
\begin{equation}\label{lang}
dx_t = \chi(x_t)F_t(x_t)dt + \nabla D(x_t)dt + \sqrt{2D(x_t)} dB_t
\end{equation}
where we shortened notation by defining
$F_t(x) = f_t(x) - \nabla U(x)$. Note that the force which the particle undergoes depends on time in two ways:
first of all because the force depends on the position of the particle and the position
depends on time, and secondly because the force changes in time according to some
deterministic periodic protocol. Note also that the system is now driven from equilibrium, 
because of the time-dependence of the dynamics.
Our analysis includes the case in which the force $f_t$ is nonconservative, meaning there is no potential from which it derives,
and as a result currents are generated.\\

\noindent The corresponding Fokker-Planck equation for distributions $\mu_t$ is now:
\begin{equation}\label{fokpla}
\frac{d\mu_t(x)}{dt} + \nabla \cdot J_{\mu_t}(x) = 0
\end{equation}
with the probability current $J_{\mu_t}$ given by
\begin{equation}
 J_{\mu_t}(x) = \chi(x)F_t(x)\mu_t(x)-D(x)\nabla\mu_t(x)
\end{equation}
As the forcing is now time-dependent, we do not expect to find a stationary distribution. But because the forcing is
periodic in time, what we expect is that given a long enough time, measures $\mu_t$ will `relax' to an
evolution that is also periodic in time. In other words we assume that there is a measure $\rho_t(x)$ that solves the Fokker-Planck
equation (\ref{fokpla}) and satisfies $\rho_t = \rho_{t+\tau}$, and moreover $\mu_t\to\rho_t$ for $t\to\infty$.
We will call this a non-equilibrium oscillatory
state (NOS). Note that such a NOS becomes a non-equilibrium steady state (NESS) when the period of the forcing
goes to infinity and the dynamics become time-independent.

\subsection{Questions}

As said before, when one considers particle densities, rather than the trajectory of one particle, the dynamics of the system is described
by a Fokker-Planck equation, which is basically a deterministic equation. The determinism arises because of the law of large numbers,
as a particle density by its very definition involves many particles. But for mesoscopic systems this assumption is not always valid,
and fluctuations from this average ``Fokker-Planck behavior'' become important. The first question that then arises is:
\begin{enumerate}
 \item[1.] \textit{What are the right observables?}
\end{enumerate}
These observables will be the empirical occupations $\mu_{n,t}$ and currents $j_{n,t}$, where $n$ stands for the number of
periods ($T = n\tau$) that have lapsed and $t\in [0,\tau]$ for a specific time in each period. Then e.g. $\mu_{n,t}(A)$ counts
the number of periods in which the particle is in a subset $A$ of the total space at this specific time $t$ within each period.
We will explain this more exactly in Section \ref{observables}.\\

\noindent What we expect is that $\mu_{n,t} \to \rho_t$ and $j_{n,t}\to J_{\rho_t}$ when we let $n\to\infty$. But, as the process is stochastic,
for large but finite numbers of periods, anything can happen. There is a probability that, even after a long time, the system has not yet
reached the NOS, but $\mu_{n,t}$ resembles some other distribution, $\mu_t$ say. It is here that we enter the domain of dynamical fluctuations
and large deviations \cite{DZ,var}. The probability of this happening will be exponentially small in time:
\[ P(\mu_{n,t} = \mu_t) \propto e^{-T I(\mu_t)} \]
where $I(\mu_t)$ is called a rate function, or a fluctuation functional. In the same way we can define $I(j_t)$ for the currents,
and $I(\mu_t,j_t)$ for the joint fluctuations of occupations and currents. It is here that the following and main questions arise:
\begin{enumerate}
 \item[2.] \textit{Can we calculate the fluctuation functionals?}
 \item[3.] \textit{What is their thermodynamical meaning?}
\end{enumerate}
Concerning question 2: in Section \ref{secoccur} we explicitly calculate $I(\mu_t,j_t)$, see (\ref{occurfluc}). From this $I(\mu_t)$ and $I(j_t)$
can in principle be calculated through variational calculus, as $I(\mu_t) = \inf_{j_t}I(\mu_t,j_t)$ and $I(j_t) = \inf_{\mu_t}I(\mu_t,j_t)$.
Explicitly calculating these is hard, but we work out some specific examples in the ensuing chapters.\\

\noindent The main purpose of this text is to discuss the thermodynamic meaning of these fluctuation functionals. This is done in
several steps at several places. First in Section \ref{secenttraf} we define and discuss the main thermodynamic ingredients
used throughout our argument: entropy and traffic. As we will see, entropy is time-antisymmetric and the less-known traffic is its time-symmetric
counterpart. The latter is a measure of the dynamical activity of the process. Then, in Section \ref{secfluc},
we keep track of these ingredients throughout the calculations. Additional thermodynamic interpretation is then provided in
Section \ref{canonical}, where we try to make a connection with the idea of thermodynamic potentials.\\

\noindent Finally, we study the limits of slow dynamics and small fluctuations.
With a slow dynamics we mean here that the time-derivative of the forcing is small. This will be discussed in Section \ref{smallslow}.

\section{Entropy and traffic}\label{secenttraf}
Our setting is a driven diffusive system which is not in equilibrium. Constantly, there is an entropy flux between system (the Brownian particle)
and environment (the heat bath). This entropy flux is of course just a heat flux. If the system has at a certain time
a density $\mu_t$, then the heat flux into the environment
is equal to minus the change in energy of the system plus the work done on the system:
\begin{equation}\label{heat}
 Q_t(\mu_t) = \int f_t(x) J_{\mu_t}(x) dx -\frac{d}{dt}\int U(x)\mu_t(x) dx
\end{equation}
Indeed, the first term on the right-hand side is the work done by the nonconservative force and the second is
the energy change.
Note further that this flux is a flux per unit of time (i.e. it is a rate).

\noindent It has been known for a while now \cite{mn} that the entropy production can be defined and computed by comparing the probabilities
of trajectories to the probabilities of time-reversed trajectories. This way of looking at entropy production is very convenient
in dynamical fluctuation theory because it makes clear why it plays such an important role in the fluctuation functionals. On the
other hand, in this way it also becomes clear that entropy is not the only important quantity. Entropy has its counterpart, called
traffic (see \cite{cmaes08a,cmaes08b,cmaes08c}), that becomes very important when going out of equilibrium. Therefore we will define and compute these
quantities in this section, before going to the actual computations of fluctuation functionals in the next section. To do this,
we first need to know how to compute probabilities of trajectories.

\subsection{Trajectories}\label{trajectories}
In statistical physics it is important to calculate expectation values of observables. These expectation values
are actually averages over all possible realizations of the stochastic process at hand. For our model such a realization
is the successive positions the Brownian particle visits during an interval $[0,T]$,
this is called a \textit{trajectory} or \textit{path}, which we denote by $\omega = (x_t)_{0\leq t\leq T}$.
The probability density of such a path is denoted by
$P_{\mu_0}(\omega)$, where $\mu_0$ denotes the density from which the initial position $x_0$ is chosen.
Expectation values of observables $\caO$ are then computed/defined by integrating over all possible trajectories:
\[ \left<\caO(\omega)\right>_{\mu_0} := \int d\omega P_{\mu_0}(\omega)\caO(\omega) = \int dP_{\mu_0}(\omega)\caO(\omega) \]
Due to the Gaussian nature of the noise in (\ref{lang}),
the probability density of observing a path $\omega$ is proportional to\cite{cmaes08a}.
\[ P_{\mu_0}(\omega) \propto \mu_0(x_0)\exp\left\{-\frac{1}{4}\int_0^Tdt[\dot{x}_t-\chi F_t-\nabla D]D^{-1}[\dot{x}_t-\chi F_t-\nabla D]\right\} \]
where we have, for the sake of clarity, notationally suppressed the dependance of all functions on $x_t$.
Actually, for the analysis in the rest of this text, we will use only relative probability densities.
In this section we discuss the relative probability density of our process with respect to the reference process given by \ref{eqref}
Denoting the path-probability density of the reference by $P^0$, we get
\begin{equation}
 \frac{dP_{\mu_0}}{dP^0_{\rho_0}}(\omega) =: \frac{\mu_0(x_0)}{\rho_0(x_0)}e^{-A(\omega)}
\end{equation}
where the \textit{action} $A$ is given by
\begin{equation}\label{action}
A(\omega) = -\frac{\beta}{2}\int_0^Tdx_t\circ f_t
+\frac{\beta}{4}\int_0^Tdt[f_t + \frac{2}{\beta}\nabla-2(\nabla U)]\chi f_t
\end{equation}
in which the circle $\circ$ denotes a Stratonovich-type stochastic integral (see \ref{apB}).
This relative density, described by the action (\ref{action}) is the key quantity needed to compute
the fluctuation functionals in our framework. We closely examine its physical
interpretation by splitting the action $A$ into its time-symmetric and time-antisymmetric parts.
We define the time-reversal operator $\theta$ as follows: it reverses the trajectories in time
$\theta\omega = (x_{T-t})_{0\leq t\leq T}$, and it also reverses the protocol of the forcing $\theta f_t = f_{T-t}$.
The anti-symmetric and symmetric parts of the action are then defined as
\begin{eqnarray}
 S(\omega) &=& \theta A(\omega) - A(\omega) = \beta\int_0^Tdx_t\circ f_t(x_t)\label{enttraf}\\
\nonumber \caT(\omega) &=& \theta A(\omega) + A(\omega) = \frac{\beta}{2}\int_0^Tdt[f_t(x_t) + \frac{2}{\beta}\nabla-2(\nabla U(x_t))]\chi(x_t) f_t(x_t)
\end{eqnarray}
Physically, $S(\omega)$ is the excess entropy flux into the environment during the trajectory $\omega$,
excess with respect to the equilibrium reference process. We will name $T(\omega)$ the \textit{traffic} in
accordance with\cite{cmaes08a}.

\subsection{Entropy}

It has been known for a while now, that entropy production is a measure of irreversibility.
In \cite{mn} it was argued that the total entropy change in the world as a consequence of the process
is given by
\[ S_{tot}(\omega) = \log \frac{dP_{\mu_0}(\omega)}{dP_{\mu_T}\theta(\omega)} \]
where $\mu_T$ is the measure that arises when $\mu_0$ is evolved through time via the Fokker-Planck equation (\ref{fokpla}).
Indeed, for an equilibrium process, like the reference process (\ref{eqref}), this will be zero:
\[ S^{eq}_{tot}(\omega) = \log \frac{dP^0_{\rho_0}(\omega)}{dP^0_{\rho_0}\theta(\omega)}=0 \]
This means that we can write
\begin{eqnarray*}
 S_{tot}(\omega) &=& \log \frac{dP_{\mu_0}(\omega)}{dP_{\rho_0}(\omega)} - \log \frac{dP_{\mu_T}\theta(\omega)}{dP_{\rho_0}\theta(\omega)}\\
&=& \log \frac{\mu_0(x_0)}{\rho_0(x_0)} - \log \frac{\mu_T(x_T)}{\rho_0(x_T)} + S(\omega)
\end{eqnarray*}
with $S(\omega)$ the antisymmetric part of the action as defined in (\ref{enttraf}).
As said before, this $S$ is the excess entropy flux into the environment.
It is the work done by the force $f_t$ times $\beta$.\\

Let us compute the expectation value of $S_{tot}$. The average of a Stratonovich integral is given in \ref{apB}, see (\ref{a9}):
\[ \left<S_{tot}\right>_{\mu} := \int dP_{\mu}(\omega)S_{tot}(\omega) = S_R(\mu_0)-S_R(\mu_T) + \beta\int_0^Tdt\int dx f_t J_{\mu_t}\]
where $S_R(\mu)$ is the relative entropy of the measure $\mu$ with respect to the reference equilibrium measure:
\[ S_R(\mu) = \int dx \mu(x)\log \frac{\mu(x)}{\rho_0(x)} \]
Using the explicit form of the reference equilibrium measure $\rho_0 \propto e^{-\beta U}$, we can rewrite $S_{tot}$ in a way
that is independent of the reference:
\[ \left<S_{tot}\right>_{\mu} := \int dP_{\mu}(\omega)S_{tot}(\omega) = s(\mu_T)-s(\mu_0) + \beta\int_0^Tdt\int dx [f_t J_{\mu_t}-\dot{\mu}_tU] \]
where $s(\mu)$ is now the Shannon entropy associated to the distribution $\mu$
\[ s(\mu) = -\int dx \mu(x)\log \mu(x) \]
Compare this to (\ref{heat}). We see that the total entropy production is the change in shannon entropy plus the entropy flux between
system and environment.\\

Using the Fokker-Planck equation we can still rewrite the total average entropy production in its shortest form:
\begin{equation}
 \left<S_{tot}\right>_{\mu_0} = \int_0^Tdt \sigma_t(\mu_t)
\end{equation}
with
\[ \sigma_t(\mu_t) := \int dx J_{\mu_t}(\mu_t D)^{-1}J_{\mu_t} \]

\subsection{Traffic and its relation to entropy}

There is still a part of the action we have not discussed yet, that is, the time-symmetric part, or the traffic, second equation in (\ref{enttraf}).
In contrast to the entropy production, this traffic depends very much on the reference process we take, so one should always keep in mind
that traffic is an excess w.r.t to this reference. The average of the traffic can be computed and will be useful:
\begin{equation}\label{avtraffic}
 \left<\caT\right>_{\mu_0} = \int_0^T dt \tau_t(\mu_t)
\end{equation}
with
\[ \tau_t(\mu_t) := \frac{\beta}{2} \mu_t[f_t -2(\nabla U)+\frac{2}{\beta}\nabla]\chi f_t \]
A straightforward calculation then connects this average traffic to the total entropy production:
\begin{equation}\label{enttrafrel}
 \tau_t(\mu_t) = \frac{1}{2}\sigma_t(\mu_t) - \frac{1}{2}\sigma_t^{f_t=0}(\mu_t)
\end{equation}
where the second term on the right-hand sign is the entropy production in the reference equilibrium dynamics, but computed
for a (nonequilibrium) distribution $\mu_t$.

\section{Dynamical fluctuations}\label{secfluc}

In this section we want to examine the fluctuation functionals that govern the asymptotic probabilities of observing
nonstationary occupations and currents. For this it is very important to correctly define the observables we want to consider.

\subsection{Defining the observables}\label{observables}

To correctly define our observables we assume that the time of measurement $T$ is very long, 
and a multiple of the period $\tau$: $T=n\tau$. We then define the empirical occupation density $\mu_{n,t}(x)$ as follows:
\begin{equation}\label{defmun}
 \mu_{n,t}(x) = \frac{1}{n}\sum_{k=0}^{n-1}\delta(x_{t+k\tau}-x)
\end{equation}
This density `counts' for every time $0\leq t\leq\tau$ the number of periods in which the 
particle is at position $x$ at this specific time within each period. In this sense this 
empirical density is very detailed. As a function of $x$ it satisfies $\int dx\mu_{n,t}(x)=1$. 
The reason for defining this detailed density is that it has the property that for any function $w_t(x_t)$
\begin{equation}\label{occdef}
 \frac{1}{T}\int_0^Tdtw_t(x_t) = \frac{1}{\tau}\int_0^{\tau}dt\int dx w_t(x)\mu_{n,t}(x)
\end{equation}
As a consequence of ergodicity, we have that $\frac{1}{T}\int_0^Tdtw_t(x_t)\to\int dx w_t(x)\rho_t(x)$ for $T\to\infty$.
This shows us that for $T\to\infty$ we have that $\mu_{n,t}\to\rho_t$ almost surely. In other words:
\begin{eqnarray*}
 \lim_{T\to \infty}\caP(\mu_{n,t}=\mu_t) &=& 0, \ \ \ \ \ \ \mu_t\neq \rho_t\\
  \lim_{T\to \infty}\caP(\mu_{n,t}=\rho_t) &=& 1
\end{eqnarray*}
However, if $\mu_t\neq \rho_t$, one can compute
\begin{equation}\label{asymptotic}
  I(\mu_t) := -\lim_{T\to \infty}\frac{1}{T}\log\left(\caP(\mu_{n,t}=\mu_t)\right)
\end{equation}
where $I(\mu_t)$ is called a rate function, or a fluctuation functional. If it is finite,
then we can schematically write
\begin{equation}
 \caP(\mu_{n,t}=\mu_t)\propto e^{-TI(\mu_t)}
\end{equation}
by which we see that the probabilities of fluctuations $\mu_t$ are exponentially damped in time.
If $I(\mu_t)=\infty$, then the probability of the fluctuation $\mu_t$ is damped even stronger.
We will only consider fluctuations for which the fluctuation functional is finite.
We have to admit that the equality sign in the probabilities $\caP(\mu_{n,t}=\mu_t)$ should not be taken too seriously,
the precise mathematical formulation of this and of rate functions can be found in\cite{DZ}. A more 
intuitive approach to large deviations is given in\cite{var,tou}.\\

\noindent The empirical occupation density is manifestly time symmetric. Its time antisymmetric counterpart is given by
\begin{equation}
 j_{n,t}(x)dt = \frac{1}{n}\sum_{k=0}^{n-1}dx_{t+k\tau}\circ\delta(x_{t+k\tau}-x)
\end{equation}
This measures the current at $x$ at a specific time $0\leq t\leq \tau$ in each period, and averages 
over $n$ periods. By this definition we see that for any function $w_t$
\begin{equation}\label{curdef}
 \frac{1}{T}\int_0^Tdx_t\circ w_t(x_t) = \frac{1}{\tau}\int_0^{\tau}dt\int dx w_t(x)j_{n,t}(x)
\end{equation}
By this and ergodicity we can also see, together with formula (\ref{a9}), that for $T\to\infty$ we get $j_{n,t}\to J_{\rho_t}$.
In the same sense as for occupations, one can define the rate function $I(j_t)$.\\

\noindent A more central and explicit starting point however is the rate function for the joint fluctuations of $\mu_t$ and $j_t$, defined by
\begin{equation}
 \caP(\mu_{n,t}=\mu_t,j_{n,t}=j_t)\propto e^{-TI(\mu_t,j_t)}
\end{equation}
An important note before we go on calculating rate functions is that $I(\mu_t,j_t) = +\infty$ whenever $\mu_t\neq \mu_{t+\tau}$ or 
$\frac{d\mu_t}{dt}+\nabla\cdot j_t\neq 0$, see \ref{apA}. Therefore we do not consider these cases.

\subsection{Joint fluctuations of occupations and currents}\label{secoccur}

To compute $I(\mu,j)$ we use a standard large deviation technique, sometimes referred to as Cram\'er tilting,
see e.g. \cite{DZ,var}. Here is the recipe: modify
 the driving of the original dynamics, changing $f_t$ into some $g_t$ and we take care that
 this new and modified dynamics is chosen  so that $\mu_t$ and $j_t$ become both
typical in the sense that
\begin{equation}
 \frac{d\mu_t}{dt} + \nabla\cdot j_t = 0 ,\ \ \ \ \ \ j_t = \chi(g_t-\nabla U)\mu_t - D\nabla \mu_t
\end{equation}
which explicitly defines $g_t$.\\

We will prove that
\begin{equation}\label{occurfluc}
 I(\mu_t,j_t)  = \frac{1}{2\tau}\int_0^{\tau}dt\tau^f_t(\mu_t) - \frac{1}{2\tau}\int_0^{\tau}dt\tau^g_t(\mu_t) +
  \frac{\be}{2\tau} \int_0^{\tau}dt\int dx(g_t - f_t) \cdot j_t
\end{equation}
The first two terms constitute an excess traffic averaged over one period of time ($\tau$). Note that from now on, we will
add superscripts to quantities defining in which dynamics they are computed, i.e. in the original dynamics
with $f$ or in the modified dynamics with $g$. The second term is the average work over one period of time,
done by the extra force $g_t-f_t$ that is added to the original dynamics. Note that with the relation between average traffic and entropy production
(\ref{enttrafrel}),
we can also write
\begin{equation}\label{occurfluc2}
 I(\mu_t,j_t)  = \frac{1}{2\tau}\int_0^{\tau}dt \sigma^f_t(\mu_t) - \frac{1}{2\tau}\int_0^{\tau}dt \sigma^g_t(\mu_t) +
  \frac{\be}{2\tau} \int_0^{\tau}dt\int dx(g_t - f_t) \cdot j_t
\end{equation}

To prove (\ref{occurfluc}), we write the probability of the fluctuations as
\begin{eqnarray*}
 P(\mu_{n,t} = \mu_t, j_{n,t}=j_t) &=& \int dP_{\mu_0}^f(\omega)\delta(\mu_{n,t} - \mu_t) \delta(j_{n,t}-j_t)\\
&=& \int dP_{\mu_0}^g(\omega)\frac{dP_{\mu_0}^f}{dP_{\mu_0}^g}(\omega)\delta(\mu_{n,t} - \mu_t) \delta(j_{n,t}-j_t)
\end{eqnarray*}
We use the formulas from section \ref{trajectories} to compute
\begin{eqnarray*}
 \log\frac{dP_{\mu_0}^f}{dP_{\mu_0}^g}(\omega) &=& A^g(\omega) - A^f(\omega)\\
&=& \frac{1}{2}[\caT^g(\omega) - \caT^f(\omega) + S^f(\omega) - S^g(\omega)]
\end{eqnarray*}
The expressions for entropy production and traffic are given in (\ref{enttraf}) and, and with (\ref{occdef}) and (\ref{curdef})
these can be written as
\begin{eqnarray*}
 \log\frac{dP^f}{dP^g}(\omega)
&=& \frac{T}{2\tau}\int_0^{\tau}[\tau^g_t(\mu_{n,t}) - \tau_t^f(\mu_{n,t}) + \beta (f_t-g_t)j_{n,t}]\\
&=& -TI(\mu_{n,t},j_{n,t})
\end{eqnarray*}
If we substitute this in the probability of the fluctuations, we get
\[ P(\mu_{n,t} = \mu_t, j_{n,t}=j_t) = e^{-TI(\mu_t,j_t)}\int dP_{\mu_0}^g(\omega)\delta(\mu_{n,t} - \mu_t) \delta(j_{n,t}=j_t) \]
The second factor on the right-hand side is proportional to one in the asymptotic limit of $T\to\infty$. This is because
the average is now computed in the modified dynamics, in which $\mu_t$ together with $j_t$ represent the NOS, i.e. the typical behavior.
This proves that the fluctuation functional is given by (\ref{occurfluc}).\\

Using the explicit expression for $\tau_t(\mu_t)$ (\ref{avtraffic}), the fluctuation functional can also be written as
\begin{equation}\label{imujex}
 I(\mu_t,j_t)  = \frac{1}{4\tau}\int_0^{\tau}dt\int dx[j_t-J_{\mu_t}](\mu_t D)^{-1}[j_t-J_{\mu_t}]
\end{equation}
Indeed, the fluctuation functional for the joint probability of occupations and currents is fully explicit. To compute the
fluctuations of only the occupations one `simply' needs to integrate out the currents. Schematically:
\[ e^{-TI(\mu_t)}=P(\mu_{n,t} = \mu_t) = \int dj_t e^{-TI(\mu_t,j_t)} \]
However, because we are considering the asymptotic limit of $T\to\infty$, we are left with a variational problem:
\begin{equation}\label{variational} I(\mu_t) = \inf_{j_t}I(\mu_t,j_t) \end{equation}
Vice versa we find the current fluctuations by taking the infimum over all $\mu_t$ of the joint fluctuation functional.

\section{Occupation fluctuations}\label{secoc}
Let us try to compute the infimum in (\ref{variational}): the Euler-Lagrange equations give
\begin{equation}\label{eulag}
  j_t = J_{\mu_t} -\mu_t\chi V_t  = \mu_t\chi[f_t-\nabla U - \nabla V_t] - D\nabla \mu_t
\end{equation}
where $V_t$ is a Lagrange multiplier, making sure that $\frac{\partial \mu_t}{\partial t} +\nabla\cdot j_t= 0$.
Unfortunately, there is no simple general solution to this equation. On the other hand, this equation
gives a nice physical interpretation: to find the fluctuation functional for the occupations, one has to
modify the original dynamics by adding a potential $V_t$. This $V_t$ has to be such that it makes $\mu_t$ typical, i.e.
\begin{equation}\label{Vequation}
  \frac{\partial \mu_t}{\partial t}=-\nabla\cdot(\mu_t\chi[f_t-\nabla U - \nabla V_t] - D\nabla \mu_t)
\end{equation}
The fluctuation functional is then given by (\ref{occurfluc}) with $g_t = f_t-\nabla V_t$:
\[ I(\mu_t)  = \frac{1}{2\tau}\int_0^{\tau}dt \tau^{f_t}(\mu_t) - \frac{1}{2\tau}\int_0^{\tau}dt \tau^{f_t-\nabla V_t}(\mu_t)
  +\frac{\be}{2\tau} \int_0^{\tau}dt\int dx \mu_t\frac{\partial V_t}{\partial t} \]
This is thus equal to the excess in traffic plus the work done by the extra potential. The excess here is an excess of the original w.r.t. the
modified process in which $\mu_t$ is typical. Equivalently:
\[ I(\mu_t) = \frac{\beta}{4\tau}\int_0^\tau dt \int dx \mu_t (\nabla V_t)\chi (\nabla V_t) \]
A downside of the results above is that
there is no general explicit expression for $I(\mu_t)$. Therefore we will consider some (simple)
examples where one can get explicit results.

\subsection{One dimension}
In the case of a one-dimensional space $\Omega$, it is possible to explicitly calculate the occupation fluctuation functional.
This is because the relation $\frac{d\mu_t}{dt} = - j'_t$ (where the prime denotes derivation with respect to $x$)
can be rewritten as
\[ j_t(x) = \int_a^x \frac{d\mu_t(y)}{dt}dy + j_t(a) \]
for some number $a \in \Omega$. This immediately gives the solution to (\ref{eulag}):
\[
V'_t(x)= (\mu_t(x)\chi(x))^{-1}\left[\int_{a}^x\frac{d\mu_t(y)}{dt}dy+j_t(a)\right] + f_t(x)-U'(x)-\frac{1}{\beta}\frac{\mu'_t(x)}{\mu_t(x)}
\]
Finally one can then determine $j_t(a)$ from boundary conditions.
For example, let us take $\Omega = \mathbb{R}$, and take $a=-\infty$. 
if the potential $U$ and the force are sufficiently confining, then $j_t(a)$ will be zero. This is for example
the case when $f_t(x)-U'(x)$ is a polynomial in $x$ with highest degree term equal to $cx^k$ with $c>0$ and $k$ even.\\
Another example is to take $\Omega$ to be the unit circle. We take $a=0$,
and find $j_t(0)$ by using the periodicity condition on a unit circle $V(0)=V(1)$:
\[
j_t(0) = -\frac{1}{\int_0^1 (\mu_t(x)\chi(x))^{-1}dx}\left(\int_0^1 f_t(x)dx + \int_0^1 dx \frac{1}{\mu_t(x)\chi(x)}
\int_0^x  dx'\frac{d\mu_t(x')}{dt}\right)
\]

\subsection{Harmonic potential}

In this example we work in $\mathbb{R}^d$. We take the potential to be harmonic: $U=k\sum_{i=1}^dx_i^2$,
and both the force $f_t$ and the diffusion matrix $D$ to be independent of $x$.
For any fixed time, this gives a detailed balanced process. It is the time-dependence of $f_t$ that
pulls the system out of equilibrium.
The corresponding Fokker-Planck equation is given by 
\[ \frac{d\mu_t}{dt} + \nabla J_{\mu_t} = 0 \ \ \ \ \ \textrm{with }\ \ \ J_{\mu_t} = \chi(f_t-kx_t)\mu_t - D\nabla \mu_t \]
An advantage of this model is that we can explicitly calculate the NOS:
\[ \rho_t = \frac{1}{Z}e^{-\frac{\beta k}{2}\sum_i(x_i-a_{i,t})^2} \]
where the function $a_t$ is given by
\[ a_t = \int_{-\infty}^t ds e^{-(t-s)k\chi}\chi f_s \]
This means that the NOS is a Gaussian with a mean that is oscillating. 
Even for this harmonic potential
one can not explicitly calculate $I(\mu_t)$ for a general $\mu_t$. However, we can calculate it
for a restricted class of distributions. Inspired by the NOS we take $\mu_t$ to be an oscillating
Gaussian:
\[ \mu_t = \frac{1}{Z}e^{-\frac{\beta k}{2}\sum_i(x_i-b_{i,t})^2} \]
with $b_t$ a periodic function in time. A straightforward calculation then gives the solution of (\ref{Vequation}):
\[ \nabla V_t = f_t-kb_t - \chi^{-1}\frac{db_t}{dt} \]
so that the occupation fluctuation functional becomes
\[ I(\mu_t) = \frac{\beta}{4\tau}\int_0^{\tau}dt[f_t-kb_t - \chi^{-1}\frac{db_t}{dt}]\chi [f_t-kb_t - \chi^{-1}\frac{db_t}{dt}] \]
Note that the current that minimizes $I(\mu_t,j_t)$ for this class of $\mu_t$ is equal to
$\frac{db_t}{dt}\mu_t$.

\section{Current and velocity fluctuations}\label{seccur}

As for the occupation fluctuations, a general explicit solution for the current fluctuations is not possible,
even not in the one-dimensional case. Therefore we will again examine some simple examples.
Before doing that, however, we
will explicitly check the fluctuation theorem\cite{ecv,GC}.
For this we use the expression (\ref{imujex})
\begin{eqnarray*}
 I(\mu_t,-j_t) - I(\mu_t,j_t)  &=& \frac{\beta}{\tau}\int_0^{\tau}dt\int dx j_t[f_t-\nabla U - \frac{\nabla \mu_t}{\mu_t}]\\
&=& \frac{\beta}{\tau}\int_0^{\tau}dt\int dx [j_tf_t+(U + \log \mu_t)\frac{\partial \mu_t}{\partial t}]
\end{eqnarray*}
In the last equality, the second term within the square brackets integrates to zero, because the integration is over
exactly one period of $\mu_t$. We are thus left with
\[ I(\mu_t,-j_t) - I(\mu_t,j_t) = \frac{\beta}{\tau}\int_0^{\tau}dt\int dx j_tf_t \]
and because the right-hand side does not depend on $\mu_t$ we also have that
\[ I(-j_t) - I(j_t) = \frac{\beta}{\tau}\int_0^{\tau}dt\int dx j_tf_t \]
By definition of the fluctuation functional, this is equivalent to
\[ \frac{P(j_{n,t}=j_t)}{P(j_{n,t}=-j_t)} = e^{-n\beta\int dt\int dx j_tf_t} \]
which is an instance of the fluctuation theorem.

\subsection{Divergenceless force}
We will consider here a dynamics in which $\nabla\cdot (\chi F_t) = 0$.
In this case we immediately see from (\ref{fokpla}) that $\rho_t = \frac{1}{|\Omega|}$ is just the
uniform distribution, regardless of the period of the force $F_t$. Note that we need to have
a space with a finite volume in this case. Therefore we take $\Omega$ to be the unit torus in $d$ dimensions,
which has volume $|\Omega| = 1$.
The corresponding NOS-current is $J_{\rho_t} = \chi F_t$, which is then of course divergenceless.
Let us therefore compute $I(j_t)$ for divergenceless currents. 
This means that we must minimize the joint fluctuation functional $I(\mu_t,j_t)$ over all 
(time-)constant distributions $\mu_t=\mu$. This variational problem is straightforwardly solved,
and one finds that $\mu=1$ is the minimizer for any divergenceless $j_t$. This results in a current
fluctuation functional that is quadratic in the current:
\[ I(j_t) = \frac{\beta}{4\tau}\int_0^{\tau}[j_t-\chi F_t]D^{-1}[j_t-\chi F_t] \]

\subsection{Velocity fluctuations for a harmonic potential}

Here we will revisit the example of a harmonic potential, that was also studied for occupation fluctuations.
While it is still difficult to find an explicit expression for the current fluctuations, it is possible to
find explicit expressions for velocity fluctuations. The velocity associated to a distribution $\mu_t$
and a current $j_t$ is $v_t = j_{t}/\mu_t$.

We saw in the harmonic example for occupation fluctuations, that the current that minimizes the 
joint fluctuation functional for gaussian distributions,
can be written as $v_t\mu_t$, with $v_t$ a factor independent of $x$. This factor is exactly
the velocity. In this example we will consider the reverse problem: what is the fluctuation functional for 
 velocities that are independent of $x$? 
For this we will first rewrite the fluctuation functional by a change in variables. The new probability
density $\tilde{\caP}$ for distributions and velocities is
 \[\tilde{\caP}(\mu_t,v_t) = \mu_t \caP(\mu_t,\mu_tv_t)  \]
But because the fluctuation functionals are defined as the logarithm of the probability
densities divided by the time $T$ in the limit for large times (as in (\ref{asymptotic})),
we have that
\[ \tilde{I}(\mu_t,v_t) = I(\mu_t,\mu_tv_t) \]
To find $\tilde{I}(v_t)$, we minimize the joint fluctuation functional over all $\mu_t$.
A straightforward, but somewhat tedious calculation gives that the minimizing distribution
is a Gaussian:
\[ \mu_t = \frac{1}{Z}e^{-\frac{\beta k}{2} \sum_i (x_i - b_{t,i})} \]
with the function $b_t$ equal to
\[ b_t =  \int_0^tv_sds + \frac{1}{\beta k \tau}\int_0^{\tau}dt \left[ D^{-1}v_t - \beta f_t +\beta k \int_0^tv_sds  \right] \]
The fluctuation functional for velocities then becomes:
\[ \tilde{I}(v_t) = \frac{1}{4\tau}\int_0^{\tau}[\chi^{-1}v_t-(f_t-kb_t)]\chi[\chi^{-1}v_t-(f_t-kb_t)]   \]

\section{A notion of thermodynamic potentials and variational principles}\label{canonical}

In equilibrium systems it is useful to consider thermodynamic potentials (like free energy), as they have a clear physical
meaning and characterize equilibrium via variational principles. Moreover, to go from one potential to another,
Legendre transforms are used. Fluctuation functionals also bring with them variational principles. First of all, and mainly, minimizing them
characterizes stationarity. This is easily seen for example for the occupations:
\[ \caP(\mu_{n,t}=\mu_t) \leq \caP(\mu_{n,t}=\rho_t)\propto 1\ \ \ \  \Rightarrow \ \ \ \  I(\mu_t)\geq I(\rho_t) = 0 \]
In the same way, the joint fluctuation functional characterizes the typical occupations and the typical currents. 
In (\ref{occurfluc}) we gave an explicit
expression for this functional. Let us analyze this from the viewpoint of thermodynamic potentials and Legendre transforms. First of all,
the first two terms in (\ref{occurfluc}) are time-averages of the traffic, or equivalently the entropy production (\ref{enttrafrel}).
Let us consider the time-averaged traffic:
\[ H(\mu_t,h_t) = \frac{1}{\beta}\int_0^{\tau}dt \tau^{h_t}(\mu_t) \]
where $h_t$ is an arbitrary force, which can be for example $f_t$ (the original dynamics of the system), or $g_t$ (the force that makes
$\mu_t$ and $j_t$ typical). This functional $H$ can be seen as a potential for the currents in the sense that the functional derivative
of it with respect to the force, gives
\[ \frac{\delta H}{\delta h_t(x)} = J^h_{\mu_t}(x) \]
where $J^h_{\mu_t}$ is the probability current in a dynamics with a force $h_t$, see (\ref{fokpla}). It is then natural to examine
the Legendre transform of $H$:
\[ G(\mu_t,j_t) =  \sup_{h_t}\left[ \int_0^{\tau}dt\int dx h_tj_t - H(\mu_t,h_t) \right] \]
The Euler-Lagrange equations to find the supremum give then
\[ j_t = J^h_{\mu_t} \]
so that $h_t = g_t$ is exactly the force needed to make the current $j_t$ typical, together with $\mu_t$:
\[ G(\mu_t,j_t) =  \int_0^{\tau}dt\int dx g_tj_t - H(\mu_t,g_t) \]
Vice versa, if we take the functional derivative of $G$, we get:
\[ \frac{\delta G}{\delta j_t(x)} = g_{t} \]
where $g_t$ is again the force that makes $\mu_t$ and $j_t$ typical. So $G$ is a potential for the forces, just
like $H$ was a potential for the currents, and by Legendre transforms we can switch between the two.\\

\noindent We can easily rewrite $I(\mu_t,j_t)$ in terms of $G$ and $H$:
\[ I(\mu_t,j_t) = \frac{\beta}{2\tau}\left[G(\mu_t,j_t) + H(\mu_t,f_t) -\int_0^\tau dt\int dx f_tj_t\right] \]
Note that $H(\mu_t,0) = 0$, so that $\frac{\beta}{2\tau}G(\mu_t,j_t)$ is exactly equal to the fluctuation functional
in the case that $f=0$. In other words:
\[ I_{f_t}(\mu_t,j_t) = I_{0}(\mu_t,j_t) + \frac{\beta}{2\tau}\left[H(\mu_t,f_t) -\int_0^\tau dt\int dx f_tj_t\right] \]
This is nice, because the left-hand side is a fluctuation functional for a nonequilibrium dynamics, while on
the right-hand side, the first term is a fluctuation functional in an equilibrium dynamics. The rest is thus
`the correction to equilibrium.'

\section{Small and slow}\label{smallslow}

\subsection{Small fluctuations}

In this section we consider the regime of small fluctuations. In this regime a quadratic approximation is made
of the fluctuation functionals, meaning that we are essentially in the regime of Gaussian fluctuations. More precisely, we take
$\mu_t=\rho_t(1+\epsilon\mu_{1,t})$ and $j_t = J_{\rho_t} + \epsilon j_{1,t}$ with $\epsilon$ a small parameter, so that $\mu_t$ and
$j_t$ are close to the stationary distribution and current.
Up to second order in $\epsilon$, the joint fluctuation functional then becomes:
\begin{eqnarray*}
 I(\mu_t,j_t)&=&\frac{\epsilon^2}{4\tau}\int dt\int dx [j_{1,t}-J_{\rho_t}\mu_{1,t}+\rho_tD\nabla \mu_{1,t}](\rho_t D)^{-1}\\
&& \ \ \ \ \ \ \ \ \ \ \ \ \ \ \ \  \ \ \ \ \ \ \ \ \ \ \ \ \cdot[j_{1,t}-J_{\rho_t}\mu_{1,t}+\rho_tD\nabla \mu_{1,t}]
\end{eqnarray*}
In this formula, one can see that the currents $j_{1,t}$ and occupations $\mu_{1,t}$ are coupled because of two reasons: first of all,
because of cross-terms in the fluctuation functional, secondly because of the relation $\frac{\partial \mu_{1,t}}{\partial t} = - \nabla\cdot j_{1,t}$.\\

\noindent The first of these reasons disappears in the regime of small driving, meaning that we will replace $f_t$ by $\epsilon f_t$.
First of all this means that $\rho_t = \rho_0(1+\epsilon \rho_{1,t})$ and $J_{\rho_t}$ will be of order $\epsilon$.
The fluctuation functional then further simplifies to
\begin{equation}\label{smallclose}
 I(\mu_t,j_t)=\frac{\epsilon^2}{4\tau}\int dt\int dx [j_{1,t}(\rho_0 D)^{-1}j_{1,t} + \rho_0 \nabla \mu_{1,t} D \nabla \mu_{1,t}]
\end{equation}
Indeed in this approximation there are no more cross-terms of currents and occupations.

\subsection{Slow dynamics and a comparison to time-independent dynamics}

Let us examine what happens if we take a dynamics which changes slowly in time. To parameterize this `slowness'
we take a small number $\epsilon >0$, and change the time-dependent force $f_t$ to $f_{\epsilon t}$. This means that
the period of the dynamics also changes to $\tau/\epsilon$. What changes in the fluctuation functionals? To see this, note that
the fluctuation functionals are time integrals over a period $\tau$, schematically:
\[ I = \frac{1}{\tau}\int_0^{\tau}dt\caI(h_t,\dot{h}_t) \]
where $\caI$ is a functional, which depends on some functions, here denoted by $h_t$, but which can be $\mu_t$, $f_t$, etc, (see the
expressions for the functionals in the previous section). Making the dynamics slow by inserting $\epsilon$, we get
\[ I = \frac{\epsilon}{\tau}\int_0^{\tau/\epsilon}dt\caI(h_{\epsilon t},\epsilon\dot{h}_{\epsilon t}) =  \frac{1}{\tau}\int_0^{\tau}dt\caI(h_t,\epsilon\dot{h}_t) \]
where in the last step we just rescaled the integration variable $t$. So effectively, making the dynamics slow is equivalent to
putting an $\epsilon$ in front of all time-derivatives in the fluctuation functionals.
As an example, consider the occupation fluctuations where we now get
\[ I(\mu_t)  = \frac{1}{2\tau}\int_0^{\tau}dt \tau^{f_t}(\mu_t) - \frac{1}{2\tau}\int_0^{\tau}dt \tau^{f_t-\nabla V_t}(\mu_t)
  +\frac{\be\epsilon}{2\tau} \int_0^{\tau}dt\int dx \mu_t\frac{\partial V_t}{\partial t} \]
In the limit $\epsilon\to 0$, a time-independent dynamics is recovered. For example, in the last formula this means that
the last term drops from the fluctuation functional. As a check, in this limit the results of \cite{cmaes08a} are recovered.

\subsection{Small, close to equilibrium and slow}\label{ss}

If we consider the situation of (\ref{smallclose}), and take also a dynamics that is slow (with the same $\epsilon$), then
the coupling between the currents and the occupations completely disappears, up to second order in $\epsilon$. As a consequence,
we can write
\[ I(\mu_t,j_t) = I(\mu_t) +I(j_t) \]
with
\[ I(\mu_t) = \frac{\epsilon^2}{4\tau}\int dt\int dx \rho_t \nabla \mu_{1,t} D \nabla \mu_{1,t} \]
and
\[ I(j_t) = \frac{\epsilon^2}{4\tau}\int dt\int dx j_{1,t}(\rho_t D)^{-1}j_{1,t} \]
Moreover, because of this uncoupling one can write down a minimum entropy production principle for the occupations
and a maximum entropy production principle for the currents, just like in the case of time-independent dynamics \cite{cmaes08a}.
On the other hand, here so many approximations have been made that one could question the relevance of this.

\section{Summary of the results}

In the calculation of the joint rate functional or fluctuation functional we have made use of the technique 
called Cramer's tilting, in which one changes the forcing such that the deviated observables become `typical.' 
The joint rate functional consists of excess traffic and work done by the excess forcing 
both averaged over one period of forcing, see (\ref{occurfluc}). For the calculation of single rate functions (i.e. only 
for currents or for occupations) in the asymptotic limit of large $T$, this problem can be put in the 
form of a variational problem. For some specific cases the 
occupation and current functionals are derived explicitly. For the current rate functional the fluctuation theorem is verified. 

Importantly, in analogy with equilibrium thermodynamic potentials (like free energy) the non-equilibrium rate 
functionals (for example occupation rate functional) give us variational principles that lead to stationarity 
or the typical behavior. 

The time averaged traffic can be seen as a potential for the currents, as the functional derivative of 
it with respect to the corresponding force gives us the probability current (as explained in section IV). 
The Legendre transform of the potential for currents (the time averaged traffic) is the potential for the 
forces, i.e., the functional derivative of it with respect to the current gives us force. Thus Legendre 
transforms can switch between the two. Importantly, the joint rate functional can also be written in 
terms of this Legendre transform pair. 

Finally, we have considered the case of small fluctuations (the regime of Gaussian fluctuations).  
The joint fluctuational functional can be expressed in terms of the small occupation and current 
deviations. Interestingly, their coupling disappears in the regime 
of a driving that is both small and slow (section \ref{ss}). As a cross check, we approximated the rate functional to the
 case of slow driving (the forcing period was extended to $\tau/\epsilon$, where $\epsilon \rightarrow 0^+$), 
and it reduces to the case of time independent forces, as expected.

\ack
N.S. would like to thank Prof. Christian Maes for inviting him to the the Instituut voor Theoretische Fysica, 
K. U. Leuven, where this work was started, and both authors thank him for many fruitful discussions.
B.W. is supported by FWO Flanders. 

\appendix

\section{Restrictions on fluctuations}\label{apA}

Here we will show that $I(\mu_t,j_t) = +\infty$ whenever $\mu_t\neq \mu_{t+\tau}$ or
$\frac{d\mu_t}{dt}+\nabla\cdot j_t\neq 0$.\\
 First of all, by definition (\ref{defmun}) we see that the difference between $\mu_t$ and $\mu_{t+\tau}$
is quite small:
\[ \mu_{n,t+\tau}(x) = \mu_{n,t}(x) + \frac{1}{n}[ \delta(x_{t+n\tau}-x) - \delta(x_{t}-x) ]  \]
which is true for any $\omega = (x_t)_{0\leq t \leq n\tau}$. This also means that for any smooth
bounded function $w_t$ we have
\begin{eqnarray*}
  \int_0^\tau dt \int dx w_t(x)[\mu_{n,t+\tau}(x) - \mu_{n,t}(x)] &=& \frac{1}{n} \int_0^\tau dt[w_t(x_{t+n\tau})-w_t(x_t)]\\
&\to& 0 \textrm{\ \ for\ \ } T\to\infty
\end{eqnarray*}
So we see that for any path, $\mu_{n,t} - \mu_{n,t+\tau}\to 0 $ in the distributional sense.
Secondly, for any smooth bounded function $w_t$ we have that
\begin{eqnarray*}
 \int_0^Tdt\int dx w_t[\frac{d\mu_{n,t}}{dt}+\nabla\cdot j_{n,t}]&=&-\frac{1}{n}\int_0^Tdx_t\circ\cdot \nabla w_t(x_t) -\frac{1}{n}\int_0^Tdt \dot{w}_t(x_t)\\
&=& \frac{w_0(x_0)-w_T(x_T)}{n}\\
&\to& 0 \textrm{\ \ for\ \ } T\to\infty
\end{eqnarray*}
So we see that for any path, $\frac{d\mu_{n,t}}{dt}+j'_{n,t}\to 0 $ in the distributional sense.\\
\noindent Al this means that fluctuations that do not satisfy $\mu_t = \mu_{t+\tau}$ and $\frac{d\mu_{n,t}}{dt}+\nabla\cdot j_{n,t}=0$
will have a probability that does not even survive in the asymptotic limit.

\section{Stochastic Integrals}\label{apB}
For the readers that are unfamiliar with the concept of stochastic integrals we give a brief description here, together with some useful formulas.
The formulas in this appendix resemble very much the formulas in the appendix of ref \cite{cmaes08a}. However, the fact that we have now time
 dependent forces complicates things a little, though not much.\\

\noindent During computations in this article, one frequently encounters integrals over specific trajectories (paths) $\omega = (x_t)_{0\leq t\leq T}$.
As for diffusions these paths are nowhere differentiable, integrals like $\int dx_t w_t(x_t)$ are not well defined if one interprets
 them in the Riemann way.
Instead there is room for different interpretations, the most common of which are the It\^o and Stratonovich interpretations.

\emph{It\^o integral.} For the It\^o interpretation, the integral
domain $[0,T]$ is split up in a set of discrete points
$0=t_0<t_1<\ldots<t_n=T$, with $\Delta t_j = t_j-t_{j-1}$, such
that $\De t \equiv \max_j\Delta t_j\rightarrow 0$ for
$n\rightarrow\infty$. It is important to note that for
 diffusions as used in this article we have that for $\Delta t_j\rightarrow
0$,
\begin{equation}\label{diffrel}
(\Delta x_j)_{ab}^2 = (x^a_{t_j} - x^a_{t_{j-1}})(x^b_{t_j} - x^b_{t_{j-1}})\rightarrow 2D_{ab}\Delta t_j
\end{equation}
where $x^a$ is the $a$-th component of the vector $x$, and $D_{ab}$ is the corresponding matrix element.
The stochastic integral interpreted in the It\^o way is then computed as
\begin{equation}\label{ito}
  \int_0^T w_t(x_t)\, \id x_t = \lim_{n\rightarrow\infty, \Delta t\rightarrow 0}
  \sum_{j=1}^n (x_{t_j} - x_{t_{j-1}})w_{t_j}(x_{t_{j-1}})
\end{equation}
Note that the function $w_t$ is a general time-dependent function.
We do however assume that the time dependence of the function is smooth.
This has as a consequence that it does not matter if we take $w_{t_{j-1}}$ or $w_{t_j}$ in the definition of the integral.
On the other hand it does matter that one takes the evaluation point $x_{t_{j-1}}$ in the definition.\\

\noindent For the It\^o integral one cannot use the normal rules of
integration. Instead one can check from (\ref{diffrel}) and (\ref{ito}) that for any function
$w$,
\begin{equation}
  \int_0^T  \nabla w_t(x_t) \cdot  dx_t
  = w_T(x_T) - w_0(x_0) -\int_0^Tdt \dot{w}_t(x_t)- \int_0^T  D\nabla\cdot \nabla  w_t(x_t)\,\id t
\end{equation}
where a dot means a derivative of the function with respect to time, and $D\nabla\cdot \nabla  w_t$ is
a shorthand notation for $\sum_{a,b} D_{ab}\frac{\partial^2w_t}{\partial x^a\partial x^b}$.

\emph{Stratonovich integral.}
The Stratonovich interpretation differs from the It\^o interpretation only in the points of evaluation of the function $w_t$.
In this case $w_t$ is evaluated in the midpoints of the time intervals:
\begin{equation}\label{strato}
\int_0^T w_t(x_t) \circ  dx_t = \lim_{n\rightarrow\infty, \Delta t\rightarrow 0}\sum_{j=1}^n (x_{t_j} - x_{t_{j-1}})w_t\bigl(\frac{x_{t_{j}}+x_{t_{j-1}}}{2}\bigr)
\end{equation}
where the symbol $\circ$ is commonly added as a notation to distinguish
between It\^o and Stratonovich interpretations. Our definition of time reversal is that $x_t\rightarrow x_{T-t}$ and $w_t\rightarrow w_{T-t}$.
With this definition, the Stratonovich integral is
time-antisymmetric. Note also that
\begin{equation}
  \int_0^T \nabla w(x_t) \circ \id x_t = w(x_T) - w(x_0) -\int_0^Tdt \dot{w}_t(x_t)
\end{equation}
so that the Stratonovich interpretation allows us to use the normal rules of integration.

\emph{Relation between It\^o and Stratonovich.} It is easily found from (\ref{ito}) and (\ref{strato}) that
\begin{equation}\label{str-relation}
  \int_0^T w_t(x_t)  \circ x_t = \int_0^T w_t(x_t)dx_t
  + \int_0^T D\nabla w_t(x_t)dt
\end{equation}

\emph{Averages of stochastic integrals.}
We can compute the average value of an It\^o integral by using that $x_t$ solves the
It\^o-stochastic equation (\ref{lang}):
\begin{eqnarray*}
 \Bigl\langle \int_0^T w_t(x_t)  dx_t \Bigr\rangle_{\mu_0}
&=& \Bigl\langle\int_0^T w_t(x_t)[\chi_t(x_t)F_t(x_t)+\nabla D(x_t)]dt\Bigr\rangle_{\mu_0}\\
 &&+ \Bigl\langle\int_0^T w_t(x_t)\sqrt{2D}dB_t\Bigr\rangle_{\mu_0}
\end{eqnarray*}
Observe that the last It\^o-integral has mean zero since the
integrand (evaluated at each mesh point $x_{t_{j-1}}$) and the
increment $B_{t_j} - B_{t_{j-1}}$ of the Brownian motion are
mutually independent, and the latter has zero mean. Hence, the
mean value of the It\^o integral is
\begin{eqnarray}
  \Bigl\langle \int_0^T w_t(x_t) dx_t \Bigr\rangle_{\mu_0}
&=& \Bigl\langle\int_0^T w_t(x_t)[\chi(x_t)F_t(x_t)+\nabla D(x_t)]dt\Bigr\rangle_{\mu_0}
\nonumber \\
  &=& \int_0^T \id t\, \int dx\mu_t w_t[\chi F_t + \nabla D]
\label{a8}
\end{eqnarray}
The second equality uses that $\mu_t$, defined by the
Fokker-Planck equation (\ref{fokpla}), is the evolved density at time
$t$ when starting from $\mu_0$ at time zero.\\
Using the relation between It\^o and Stratonovich, we also get:
\begin{eqnarray}
  \Bigl\langle \int_0^T w_t(x_t) \circ dx_t \Bigr\rangle_{\mu_0}
&=& \Bigl\langle\int_0^T [w_t\chi_tF_t+ w_t\nabla D + D\nabla w_t]dt\Bigr\rangle_{\mu_0}
\nonumber\\
  &=& \int_0^T \id t\, \int dx w_tJ_{\mu_t}\label{a9}
\end{eqnarray}


\end{document}